\documentstyle[prb,twocolumn,aps]{revtex}
\input psfig
\begin{document}
\draft
\preprint{\small{Applied Physics Report 95- }}
\title{
Coulomb Blockade-Effects on Quantization of Charge and Persistent Current in
a Luttinger-Liquid Ring}
\author{I. V. Krive$^{(1,2)}$, P. Sandstr{\"o}m$^{(1)}$, R. I.
Shekhter$^{(1)}$,
and M. Jonson$^{(1)}$}
\address{$^{(1)}$Department of Applied Physics, 
Chalmers University of Technology and G{\"o}teborg University, 
S-412 96 G{\"o}teborg, Sweden\\
and $^{(2)}$B. Verkin Institute for Low Temperature Physics and Engineering,
Academy of Sciences of Ukraine, 310164 Kharkov, Ukraine}
\maketitle
\begin{abstract}
We show that in a ring-shaped Luttinger Liquid (LL) in contact with an electron
reservoir --- the chemical potential in the ring being controlled by a gate
voltage $V_g$ --- both the average ring charge and the persistent current in
the ring are step-like functions of $V_g$ at low temperatures.
The step positions are determined by
the LL parameter $\alpha=v_F/s$, which therefore can be directly measured. We
study electrons both with and without spin, taking into account long-range
interactions in the ring.
\end{abstract}
\section{Introduction}
Quantum transport of charge in mesoscopic systems has been intensively
studied --- both theoretically and experimentally --- during
the last decade. Two kinds of oscillatory phenomena have frequently
been in the focus of interest:
\newcounter{beans}
\begin{list}{\roman{beans}.}{\usecounter{beans}}
\item
    Coulomb blockade of tunneling through small
    capacitance metallic grains and the oscillatory behavior of the
				conductance upon change of gate voltage
    (for a review see Ref.~\onlinecite{Likharev})
\item
    Aharonov-Bohm oscillations of thermodynamic (persistent current)
    and transport (conductance) properties of mesoscopic rings in
    weak magnetic fields (see, e.g., the review articles
    Refs.~\onlinecite{Washburn,Zvyagin}).
\end{list}
Recently, both these effects were observed in semiconductor
nanostructures. \cite{Kastner,Mailly}
Unlike in the case of metals, semiconductor
heterostructures
enable one to study the properties of electron systems of reduced
(2D and 1D) dimensionality.
In nanostructures created by laterally constraining the motion of the 2D
electron gas (2DEG) formed in an $AlGaAs/GaAs$ heterostructure, both the
characteristics of the system under measurement
(electron density, geometric form and size)
and the the nature of the leads connected to it can be controlled. This is
achieved by varying (i) the voltage, $V_{g}$, on a gate
that couples capacitively to the net charge in the device, and (ii) by
adjusting
voltages on additional (split gate) electrodes that are needed for creating the
lateral confinement. Hence, one is able to vary the parameters of the
electronic
device at hand and its interaction with electron reservoirs over a wide range.
\cite{Timp}

In  systems of reduced dimensionality electron-electron correlations
play a significant role and the naive picture of noninteracting (or weakly
interacting) electrons moving in an external electrostatic potential does not
apply. In particular, under conditions of strong transverse confinement,
when the charge motion in a nominally 2D quantum channel becomes essentially
one-dimensional,
one can expect to be able to realize the Luttinger liquid (LL) regime.
Experimental studies of transport properties of laterally confined 2DEGs
do not provide unambiguous proof of LL-like behavior of charges
in quantum wires. \cite{note1}
Therefore it seems interesting and important to examine thermodynamic
properties of quantum dots and wires and to suggest experiments which
could reveal a non-Fermi-liquid character of the electron dynamics in
systems of reduced dimensionality.

\begin{figure}
\centerline{\psfig{figure=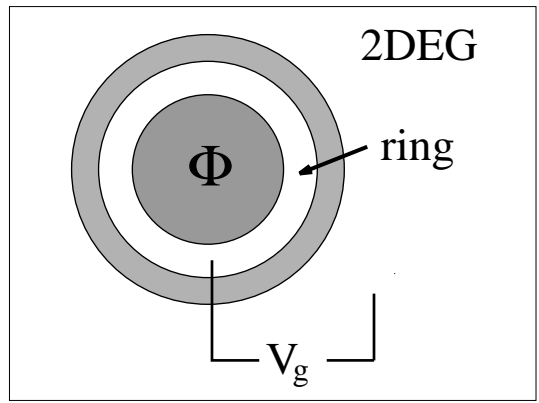,width=7cm}}
\vspace*{10mm}
\caption{ \protect{\label{fig:geometry}}
Sketch of the ring-shaped conductor discussed in the text. It encloses a
magnetic flux $\Phi$ and is created in the 2D electron gas of a gated
semiconductor heterostructure. The ring is in weak contact with an electron
reservoir, which implies that the number of electrons in the ring is not fixed
at
finite temperatures. The difference in electrostatic potential, $V_g$, between
the ring-shaped conducting area and the rest of the 2D electron gas (the
reservoir) is produced by a charged gate electrod. Because of the circular
symmetry no electric field  appears along the circumference of the ring. }
\end{figure}

As a possible test of non-Fermi-liquid behavior of 1D electrons we
propose to study persistent current oscillations and peculiarities
of charge quantization in a quantum ring controlled by a gate voltage.
The ring is assumed to be in contact with a reservoir of 2D electrons as
illustrated in Fig.~\ref{fig:geometry}. Therefore the number of ring particles
is not fixed at finite temperatures and interactions do affect transport
properties even in an impurity-free ring.
The voltage shift, $V_g$, between the conducting ring and the
rest of the 2D electron gas is produced by a charged gate electrod. In
contrast to the situation dicussed in Ref.~\onlinecite{buttiker.ps}, where
the connection between ring and electron reservoir is asymmetric,
the exchange  of electrons between ring and reservoir in our case takes place
along the entire ring circumference. Hence, it is symmetric and does not lead
to
any electric field  along the circumference of the ring.

To calculate persistent currents and
Coulomb blockade effects in a ring of strongly correlated 1D electrons we
use below the Luttinger liquid model \cite{Haldane} both for
spinless and spin-1/2 fermions.

We first show that the persistent current for electrons without spin
oscillates as a function of gate voltage,
$V_{g}$, the ring periodically
exhibiting a dia- or paramagnetic response to the
magnetic flux. The properties of these oscillations are sensitive to the
value of the correlation parameter $\alpha = v_{F}/s$ of the LL ($v_{F}$ is the
Fermi velocity, $s$ is the plasmon velocity). At low temperatures, when the
amplitude of the persistent current attains its maximum  value $I_{0}$ ($I_{0}
= ev_{F}/L$ for an impurity-free ring, $L$ is the ring circumference) the
oscillations for $\alpha < 1$ (repulsive interaction) have the form of
periodically repeated current bumps and dips. The width, i.e. the range in
gate voltage of the bumps depends on
the parameter $\alpha$ and the dimensionless flux $\Phi/\Phi_{0}$ ($\Phi_{0} =
hc/e$ is the flux quantum). In particular, at small flux values,
($\Phi\rightarrow 0$), the plateau width
for paramagnetic response is maximal for
strong repulsion ($\alpha \ll 1$) and reduces to narrow spikes in the limit of
noninteracting particles ($\alpha \rightarrow 1$). For an attractive
interaction
($\alpha > 1$) the low-$T$ oscillations of persistent current with gate voltage
disappear.

The above picture of current oscillations is correlated with the behavior
of the average ring charge, $Q$, when the gate voltage is varied. For a
repulsive interaction ($\alpha < 1$) and low temperatures, the total charge in
the ring is quantized in units of the elementary charge $e$. The function
$Q=Q(V_{g})$ exhibits a typical behavior (Coulomb "staircase"), which
is a distinct signature of the Coulomb blockade. The step positions
and their widths (for even or odd number of particles in the ring) depend on
the
correlation parameter $\alpha$. For a stiff system ($\alpha \ll 1$) all steps
are
of equal width. The width of the steps corresponding to an even number of
spinless electrons is decreased when $\alpha$ is increased and
collapses to a point when $\alpha \rightarrow 1$. For an attractive interaction
between spinless particles the mean ring charge is quantized in units of $2e$
and
the oscillations as a function of gate voltage vanish.

At high temperatures, i.e. when $T>T^{*}_{j,c}(\alpha)$ --- $T^{*}_{j,c}$ being
the crossover temperature for current ($j$) or charge ($c$) oscillations ---
the oscillations of both persistent current and average charge in
the ring become sinusoidal with an exponentially
small amplitude, $\sim \exp(-T/T^{*})$. For current oscillations the
crossover  temperature is increased if the strength of the (repulsive)
interaction is increased. In the limiting case of a stiff Wigner crystal
($\alpha \ll 1$) it has a maximum  value $T^{*} = \Delta/\pi^2$
(for spinless
electrons; $\Delta=2\pi\hbar v_F/L$ is the energy level spacing at the Fermi
level of  non-interacting electrons at the relevant density) which coincides
with the crossover temperature for an isolated LL-ring. \cite{Loss.prl,Krive}
Note that now and below $v_F$ always refer to the Fermi velocity of spinless
electrons. The actual Fermi velocity for electrons with spin and the same
density is therefore $v_F/2$.
For the charge oscillations (as function of $V_{g}$) in the limit considered
($\alpha \rightarrow 0$) the crossover temperature goes to infinity as
$T^{*}_{c}\sim 1/\alpha^2 $. This means that at any finite temperature  a ring
of strongly repulsive electrons exhibits sharp charge quantization.  In the
high-$T$ region with strong attraction, $\alpha^2 > 3 $, we predict also that
the period of the dominant contribution to the  current
oscillations is reduced by a factor of two.

The above results correspond to the case of an interaction which is
short-ranged.
We show further that long-range Coulomb forces can be incorporated into our
model
if we replace the correlation parameter $\alpha$ of the LL by the effective
"coupling"
$ \alpha_{eff}(L) = \alpha / \left[1+(2\alpha^2/\pi)(e^2/
\hbar v_F) \ln(L/\lambda)\right]^{1/2} $,
where $\lambda$ is the width of 1D channel and $L$ is the ring
circumference ($\lambda \ll L$).

The magnetic field normally used in Aharonov-Bohm experiments
is weak and the ring electrons are therefore unpolarized even at low
temperatures. To describe a real situation involving interacting
fermions with spin
one needs to understand how the electron spin affects the oscillation phenomena
of interest here. We take the spin degrees of freedom into consideration by
using the two-component LL model (spin-``up" and spin-``down"
electrons of equal
density).  Our calculation scheme can be easily adopted to this model.
For the purpose of comparing with the case of spinless electrons it is
useful to distinguish between the cases with
different parity of the
number of spin-``up" ($N_{up}^0$) and spin-``down" ($N_{down}^0$)
electrons (referring to the situation at zero temperature when the particle
number is fixed). For a repulsive interaction there are three different cases:
(i) (odd,odd) $N^0_{up}=N^0_{down} =2n+1$;
(ii) (even,even) $N^0_{up}=N^0_{down}
=2n$; (iii) (even,odd) or (odd,even) for $N^0_{up}=2n, N^0_{down}=2n+1$ or vice
versa.

We recall once more that we are dealing with a ring, which is weakly
connected to an electron reservoir of fixed chemical potential. Therefore, for
non-interacting electrons or for an attractive electron-electron interaction
between electrons carrying spin, the ring at arbitrary flux will exchange
{\em pairs}
of particles with the reservoir as the gate voltage is varied. For special
values
of the flux, when the energy levels are fourfold degenerate
($\Phi/\Phi_{0}$ = integer or half-integer), this exchange will involve four
electrons. Since for noninteracting electrons with spin an energy level is
at least doubly occupied, the
ground state ($T=0, V_{g}=0, \Phi=0$) in the ring must have an odd number of
both spin-``up'' and spin-``down'' electrons. We refer to this as the *
(odd,odd)-case, where obviously
$N_{tot}=N^0_{up}+N^0_{down}=4n+2$. The formulas for the (even,even)- case
($N_{tot}=4n$) can be derived from those valid for the (odd,odd)-case by a
simple
shift of the flux by half a flux quantum (it is the parity rule for electrons
with spin). The (odd,even)-case describes an odd number of spin-1/2 fermions in
a ring ($N_{tot}=4n+1$ or $4n+3$).
Such a ground state can be justified only for a
repulsive interaction. Moreover, for  weakly interacting particles ($1-\alpha
\ll 1$) it is destroyed due to the particle exchange with electron reservoir
already at very low temperatures $T_d \sim (1-\alpha)\Delta$.

For the (odd,odd)-case the persistent current as a function of flux behaves
very much like the current of spinless electrons. It oscillates with the
fundamental period $\Phi_0$ and amplitude $I_0$. At low temperatures
the only difference is the change in oscillation period when the gate
voltage is varied. In the limit of strong repulsion the period of
oscillations
%
%
is twice of that for spinless electrons.
The inclusion of spin changes also the high-$T$ behavior of persistent
current. The most striking effect is the decrease by a factor of three
(compared to the spinless case) of the crossover temperature in
the limit of strong repulsion.

A ring with an odd number of spin-1/2 fermions ($N_{tot}=4n+1$ or $4n+3$)
at fixed chemical potential can be conceived only at low temperatures.
In such a situation we find that the persistent current oscillates with
half the period ($\Phi_0/2$) and half the amplitude
($I_0/2$) compared with the case of spinless electrons. Notice that
for non-interacting electrons the effect of period- and amplitude ``halving"
was predicted in Ref.~\onlinecite{Loss.prb}.
We generalize this prediction to the more realistic case of correlated
electrons at finite (though small) temperatures.

The Coulomb staircase for strongly interacting spin-1/2 fermions ($\alpha
\rightarrow 0$) at low temperatures looks like the one for spinless
particles. The charge in a ring is quantized in units of the elementary
charge, $e$, and all steps are of equal width. The spin degrees of freedom
influence the Coulomb blockade effects at moderate and weak interaction
strengths. The width of steps corresponding to the number $N_{tot} = 4n, 4n+1,
4n+3$ of electrons in a ring is decreased when the interaction
strength is decreased and is contracted to a point in the limit $\alpha
\rightarrow 1$. For non-interacting spin-1/2 electrons the total charge in a
ring
at small flux values is quantized in units of $4e$ (as it should be from the
qualitative considerations).

\section{The model}

In order to study Coulomb blockade effects in a 1D ring of strongly correlated
spinless electrons, we will use the Luttinger liquid (LL) model. \cite{Haldane}
For our purposes it is convenient to represent the Lagrangian of the model in
the form:
\begin{equation}
{\cal L} = \frac{m_{0}}{8\pi^2 \overline{\rho}} \left\{ \dot \varphi^2 -
s^2(\varphi^\prime)^2
\right\} + \frac{\hbar}{L} \dot \varphi
\left( \frac{\Phi}{\Phi_{0}} + \frac{N-1}{2} \right) + \frac{\mu_{g}}{2 \pi}
\varphi^\prime	 .
\label{Lagrangian}
\end{equation}
Here $m_{0}$ is the electron (band) mass, $\overline{\rho} = N_{0}/L$ is the
mean density of
``vacuum'' electrons in the ring. [We assume that the LL-ring is in weak
contact with a reservoir of 2D electrons. Then for repulsive interaction,
$N_{0}$
is fixed by the chemical potential $\mu$ of the reservoir which is implicit
in the formulation Eq.~(\ref{Lagrangian})]; s is the plasmon velocity,
$\Phi$ is the flux of magnetic field ($\Phi_{0} = hc/e$ is the
flux quantum), $N$ is the total number of particles in the ring,
$\mu_{g}=eV_{g}$ ($V_{g}$ being the gate voltage).
The Lagrangian (\ref{Lagrangian}) describes the long-wavelength
dynamics of the electron displacement field u(x,t)
($\varphi = 2 \pi \overline{\rho} u(x,t) $). The total number, $N$,
of electrons in the ring can be represented as a sum
of ``vacuum electron'' number, $N_{0}$, and the number, $m$, of extra
electrons which appear due to the particle exchange between the
reservoir and the ring,
$N = N_{0} + m$. The integers $m = 0$,$\pm$ 1, $\pm$ 2,...) can be
connected with a winding number of topological excitations of the
$\varphi$-field in a ring geometry. It enters into the boundary conditions
imposed on the dynamical field
$\varphi (x,\tau)$ in (x,$\tau$) space ($\tau$ is an imaginary time)

\begin{equation}
\begin{array}{l}
\varphi(\tau + \beta,x) - \varphi(\tau,x) = 2 \pi n \\
\varphi(\tau,x+L) - \varphi(\tau,x) = 2 \pi m .
\end{array}
\label{boundary}
\end{equation}
Here $\beta \equiv 1/ T$  is the inverse temperature, $n = 0$, $\pm$ 1, $\pm$
2, \ldots  is the winding number for the imaginary time ``evolution''. The
twisted boundary  conditions (\ref{boundary}) are analogous to the ones
derived in  Ref.~\onlinecite{Loss.prl}.
Notice that in Haldane's description \cite{Haldane} of the LL,
the homotopy indices ($n,m$) represent topological current and
charge excitations. At nonzero flux $\Phi$, the parity rule
(see Ref.~\onlinecite{Wen})
that connects the two sectors of the model (charge and current) can be
explicitly incorporated into the Lagrangian (\ref{Lagrangian})
by adding to the external flux, $\Phi$, a fictitious (statistical) flux
$\Phi_{st} = 0$ ($\Phi_{0}/2$) for an odd (even)
total number of electrons in the
ring. This statistical flux is represented by the second term in the
parenthesis
in Eq.~(\ref{Lagrangian}).

For the following analysis, it is convenient to represent the thermodynamic
potential, $\Omega$, (grand canonical ensemble) as a path integral over
the fluctuations of the dynamical field $\varphi (x,\tau)$

\begin{equation}
\Omega = - T \ln\left\{ \sum_{m,n=-\infty}^{\infty}  \int {\cal
D}\varphi_{m,n} {\rm e}^{-S_E(\varphi_{m,n})/\hbar} \right\},
\label{potential}
\end{equation}
where $S_{e}$ is the Euclidean (imaginary time) action for the Lagrangian
(\ref{Lagrangian}) and we explicitly include in the definition,
a sum over the homotopy indices ($m,n$), which represent
the contributions of topological excitations.
%
%
%
plasmon modes.
%
%

Topological trajectories (zero modes)
satisfying the twisted boundary conditions (\ref{boundary}) take the form
\begin{equation}
\varphi_{m,n}(x,\tau) = 2\pi \frac{\tau}{\beta}n + 2\pi \frac{x}{L}m .
\label{zeromodes}
\end{equation}
A more general solution can be obtained by adding to a zero mode of fixed
$m$ and $n$ an additional part which then must have periodic boundary
conditions. This physically corresponds to having an intermediate state
that contains plasmons. Because of the periodic boundary conditions these
extra parts only depend on temperature. Since we are interested in derivates
of the thermodynamic potential with respect to flux
(to get the persistent current) and gate voltage (to get the average charge)
we will consider only the zero mode solutions in what follows. \cite{note3}

So, since only zero modes are important,
our problem reduces to the evaluation of a double sum in
Eq.~(\ref{potential}). Since the current and charge sectors of the model are
connected only by a single (parity) term in Eq.(\ref{Lagrangian}), this
summation can be done exactly. \cite{note2} The result is

\begin{eqnarray}
\label{Twoomegas}
\Omega - \Omega_{pl}&&= -T \ln\left\{ (\frac{4\pi}{\beta
{\Delta}})^{1/2} \exp(\alpha^{2}\frac{\mu^{2}_{g}
\beta}{{\Delta}})\right. \times \\ && \left.
\left[ \theta_{3}(\Theta,q) \theta_{3}(\overline{\mu},q^{\alpha^{2}}) +
\theta_{4}(\Theta,q) \theta_{4}(\overline{\mu},q^{\alpha^{2}}) \right]
\right\}.
\nonumber
\end{eqnarray}
Here $\theta_{3,4}(\nu,q)$ are the Jacobi theta-functions
(see e.g. Ref.~\onlinecite{Bateman}) and
\begin{equation}
\Theta \equiv \frac{\Phi}{\Phi_{0}} + \frac{N_{0}-1}{2} , \quad
\overline{\mu} \equiv \alpha^{2} \frac{\mu_{g}}{{\Delta}} , \quad
q \equiv \exp(-\pi^2 \frac{T}{\Delta}).
\label{six}
\end{equation}
The dimensionless parameter $\alpha = v_{F}/s$, where
$v_{F} = \pi \hbar N_{0} / m_{0}L$, characterizes the correlation properties of
the LL.

\section{Persistent current}

The persistent current in a 1D LL ring and its dependence on enclosed magnetic
flux and gate voltage can be obtained from the thermodynamic potential
(\ref{Twoomegas}). One finds that

\begin{eqnarray}
\label{current1}
I(\Phi,&&V_{g},T)= \frac{eT}{\hbar} \frac{1}{2 \pi} \times \\
&& \frac{d}{d \Theta}\ln
\left\{ \theta_{3}(\Theta,q) \theta_{3}(\overline{\mu},q^{\alpha^{2}}) +
\theta_{4}(\Theta,q) \theta_{4}(\overline{\mu},q^{\alpha^{2}}) \right\}.
\nonumber
\end{eqnarray}
At first we consider the case of free spinless electrons ($\alpha$=1). In this
case  Eq.~(\ref{current1}) can be shown to coincide with a well-known
expression for the persistent current of noninteracting electrons at fixed
chemical potential that was derived in a standard approach
using the Fermi-Dirac
distribution function. \cite{Kulik70,Cheung}
In order to show this equivalence we take advantage of the fact that for
$\alpha$=1 the sum of $\theta$-functions in Eq.~(\ref{current1}) can be
written as a product of $\theta$-functions with the help of addition formulae
for $\theta$-functions. \cite{Wittaker} The derivate with respect to flux
can then be performed to give the result
\begin{eqnarray}
\label{current2}
I_{free}(\Phi,V_{g},T)&=& \frac{eT}{\hbar} \frac{1}{4 \pi} \times \\
&&\left\{
\frac{ \theta_{3}^{'}(+)} {\theta_{3}(+)} +
\frac{ \theta_{3}^{'}(-)} {\theta_{3}(-)} +
\frac{ \theta_{4}^{'}(+)} {\theta_{4}(+)} +
\frac{ \theta_{4}^{'}(-)} {\theta_{4}(-)} \right\} ,
\nonumber
\end{eqnarray}
valid in the limit $\alpha=1$,
In Eq.~(\ref{current2}) $\theta^{'}(\nu,q)$ is the derivative of the
theta-function with  respect to its argument $\nu$, and we have used
the reduced  notation
\begin{equation}
\theta_{j}(\pm) \equiv \theta_{j}
\left( \frac{\Theta \pm (\overline{\mu}-\frac{1}{2})} {2} ,q \right).
\end{equation}
By making use of the formulae for logarithmic derivatives of the
$\theta$-functions \cite{Bateman} one finds

\begin{equation}
I_{free} = \frac{eT}{\hbar} \sum_{n=1}^{\infty} (-1)^{N_{0} n}
 \frac{ \sin(2 \pi n \frac{\Phi}{\Phi_{0}})
\cos(2 \pi n \frac{eV_{g}}{{\Delta}}) }
{ \sinh(2\pi^2 n \frac{T}{\Delta})} .
\label{current3}
\end{equation}
This expresion coincides with the well-known expression for the
persistent  current
of free spinless electrons at fixed chemical potential
$\mu$ ($\mu = \epsilon_{F} + eV_{g}$, $\epsilon_{F}$ is the Fermi energy).
Notice that the crossover temperature
$T^{*} = \Delta/2 \pi^2 = \hbar v_{F}  / \pi L$
in Eq.~(\ref{current3}) is reduced
by a factor 2 in comparison with the one for an isolated ring (fixed number of
particles). In what follows we will show that for interacting fermions
($\alpha \neq$ 1) the crossover temperature --- for a ring at fixed chemical
potential --- smoothly interpolates between the result for a stiff Wigner
crystal
(when fluctuations of the electron number are strongly suppressed) and the
result for a Fermi gas of spinless non-interacting electrons.

\begin{figure}
\centerline{\psfig{figure=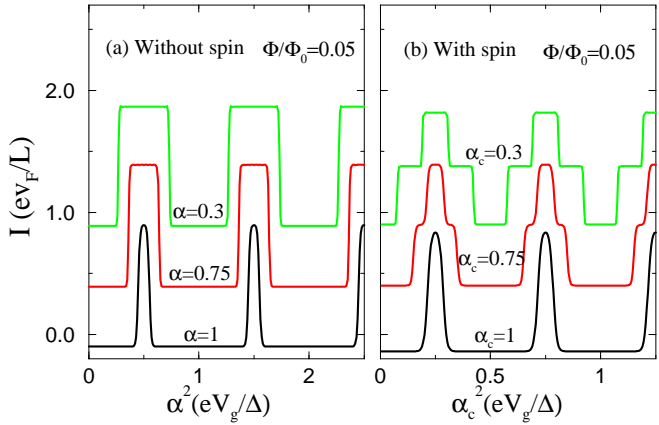,width=10cm}}
\vspace*{5mm}
\caption{ \protect{\label{fig:IofV}}
Gate voltage dependence of the persistent current for a system of electrons
without (a) and with (b) spin. Traces have been vertically offset for clarity.
In both cases the electron density is the same
and the gate voltage has been scaled by $\Delta$,
the level spacing at the
Fermi level in the spinless case.
If $\alpha=\alpha_c=1$ (free electrons) one can see a doubling of the
oscillation period in the case without spin compared to with spin.
This is because if the densities are the same,
the actual level spacing at the Fermi level
in the spinless case is twice as large as in the case with spin.
In contrast, for the case of strong interaction ($\alpha$, $\alpha_c$ small)
the ratio $\alpha_c/\alpha\to 1/2$
and hence the oscillation period for electrons with spin is now half the
period found for the case of electrons without spin. [An explicit expression
for $\alpha$ in terms of the interaction potential $V_0$ can be derived along
the lines of section V. The result $\alpha=[1+V_0/\pi\hbar v_F]^{-1/2}$
differs from Eq.~(32) for the case of electrons with spin only because the
Fermi velocity is twice as large for spinless electrons of the same density].
}
\end{figure}

Now we proceed to the case of interacting fermions.
We will use the general result (\ref{current1}) to find analytic expressions
for
the low- and high temperature limits and rely on numerical
evaluations of (\ref{current1}) in the intermediate temperature region. At low
temperatures, oscillations of the persistent current as a function of gate
voltage  (for $\Phi \rightarrow$ 0) are shown in Fig.~\ref{fig:IofV}a.
The traces shown refer to a ring with a fixed number, $N_{0}$, of ``vacuum''
fermions. It is easy  to see from the general expression Eq.~(\ref{Twoomegas})
that the result for  an ``even'' ground state can be obtained from those
corresponding to $N_{0}$ being an odd integer by a shift of the gate voltage,
$V_{g} \rightarrow V_{g} + {\Delta}/2e\alpha^{2}$. Therefore in what
follows we will analyze only the case when the number of ``vacuum'' electrons
is odd.

The oscillations of the persistent current when the
gate voltage is changed in the presence of a small magnetic flux amount to a
series of transitions between a dia- and a  paramagnetic response.
At low temperatures
these transitions have the form of periodic rectangular bumps as shown in
Fig.~\ref{fig:IofV}(a).
The width in gate voltage of the bumps and their positions depend on the
correlation parameter $\alpha$ of the LL and the dimensionless flux $\Phi/
\Phi_{0}$. The ratio of the width of paramagnetic response to the
width of diamagnetic response can be simply expressed as

\begin{equation}
\frac{\Delta V_{g}^{(p)}} {\Delta V_{g}^{(d)}} =
\frac{ \frac{1}{\alpha^{2}}-1+4 \frac{\Phi}{\Phi_{0}} }
     { \frac{1}{\alpha^{2}}+1-4 \frac{\Phi}{\Phi_{0}} } .
\end{equation}

For a stiff Wigner crystal-ring ($\alpha \ll$ 1) the width of the bumps
and the dips are equal since the ground state of a strongly repulsive
LL-ring does not depend on the parity of the particle number.
With less stiffness (larger $\alpha$), the width of the
current bumps (corresponding to a paramagnetic response) decreases and for
$\alpha \rightarrow$ 1 they reduce to narrow spikes located at $V_{g}^{(p)} =
(2k+1){\Delta}/2$, where $k$ is an integer. For an attractive
electron-electron interaction ($\alpha > 1$) , the oscillatory dependence on
gate voltage vanish. The most plausible values of the
correlation parameter should be  $\alpha \alt 1$ (we have in mind possible
experiments in AlGaAs  heterostructures). In this regime the shape of the
current oscillations is sensitive to $\alpha$ and the mere detection of such
oscillations could be a strong argument in favour of a non-Fermi liquid
like behavior of quantum rings.

In the high temperature regime $T \agt T_{j}^{*}(\alpha)$,
where $T_{j}^{*}(\alpha)$ is a crossover temperature to be specified below,
the flux dependence of the persistent current can
be expressed as a sum of two basic harmonics

\begin{eqnarray}
\nonumber
I(\Phi)&\simeq & 4 \frac{eT}{\hbar} \left\{ (-1)^{N_{0}}
{\rm e}^{- \pi^2(1+\alpha^{2})T/\Delta}
\cos(2\pi\alpha^{2} \frac{e V_{g}}{\Delta})  \times
\right. \\
&& \left. \sin(2 \pi \frac{\Phi}{\Phi_{0}}) -
{\rm e}^{-4 \pi^2 T/\Delta} \sin(4 \pi \frac{\Phi}{\Phi_{0}}) \right\}
\label{twoharm}
\end{eqnarray}
According to Eq.~(\ref{twoharm}), the main contribution
to the current is given by the first term if $\alpha^{2} < 3$. The current
then oscillates as a function of flux with the fundamental period $\Phi_{0} =
hc/e$. At the same time it is parity sensitive and has an
oscillatory dependence
on the gate voltage. From Eq.~(\ref{twoharm}) one finds the crossover
temperature
to be a function of the interaction parameter $\alpha$:

\begin{equation}
T_{j}^{*}(\alpha) = \frac{1}{\pi^2} \frac{\Delta}{1+\alpha^{2}}
\label{crossover}
\end{equation}
This temperature attains its maximum ($T^{*} = \Delta / \pi^2$) for a stiff
($\alpha \ll$ 1) Wigner crystal-ring, when the fluctuations of particle number
in the ring are suppressed and the system can be considered as
isolated. \cite{Loss.prl,Krive}

If $\alpha^{2} >$ 3, on the other hand, the second harmonic
in Eq.~(\ref{twoharm}) gives the main
contribution to the  persistent current. There is no  dependence on gate
voltage and no parity effects in this term. The current is periodic in flux
with a halved period ($\Phi_0/2$) and is diamagnetic. The
crossover temperature saturates at  $T_{j}^{*}(\alpha = \sqrt{3}) =
\Delta/4\pi^2$, which is the lowest  crossover temperature for spinless
particles. The current oscillations at $\alpha > \sqrt{3}$ resemble the
Aharonov-Bohm oscillations  in a superconducting ring (period halving, absence
of parity effect). \cite{Buttiker86}
This analogy appears also when studying the effects of
Coulomb blockade in quantum rings.

\section{Charge Oscillations and Coulomb blockade}
The average number of particles in a ring connected to a reservoir of electrons
with a chemical potential $\mu$ can be found by making use of the general
thermodynamic relation

\begin{equation}
\overline{N} = N_{0}(\mu) - \frac{1}{e} \frac{\partial \Omega}{\partial V_{g}}.
\label{meanN}
\end{equation}

Since we have an  exact expression (\ref{Twoomegas}) for the thermodynamic
potential $\Omega$ of a quantum ring, it is straightforward to find out how
$\Delta N = \overline{N} - N_{0}(\mu)$ depends on temperature, magnetic field
(flux) and gate voltage.

At low temperatures the gate voltage-dependence of charge
$Q=e\Delta N = ef(V_{g})$ takes the form of a
Coulomb staircase \cite{Kulik}
 with steps of even and odd heights (in units of $e$)
and different widths (Fig.~\ref{fig:QofV.nospin}a,b).
Using Eqs.~(\ref{Twoomegas}) and (\ref{meanN}) one can derive an analytic
expression for $\Delta N(V_{g})$. In the limit of zero magnetic flux one finds

\begin{eqnarray}
\label{deltaN}
\Delta N(\overline{\mu})&=& 2\overline{\mu} + \times \\
&&\frac{2}{\pi} \left\{
\begin{array}{ll}
\sum_{n=1}^{\infty} (-1)^{n} \frac{\sin(2\pi n\overline{\mu})}{n} &
, \frac{2k-1}{2} \leq \overline{\mu} < k+\overline{\mu}_{c} \\
\sum_{n=1}^{\infty}\frac{\sin(2\pi n\overline{\mu})}{n} &
, k=\overline{\mu}_{c} \leq \mu \leq \frac{2k+1}{2}
\end{array}
\right.
\nonumber
\end{eqnarray}
where $k$ is an integer and
\begin{equation}
\overline{\mu} = \alpha^{2} \frac{eV_{g}}{{\Delta}} \quad ,
\overline{\mu}_{c} = \left\{
\begin{array}{ll}
\frac{1}{4} (1+\alpha^{2}) & , \alpha \leq 1 \\
\frac{1}{2} & , \alpha > 1 .
\end{array}
\right.
\end{equation}
According to Eq.~({\ref{deltaN}) the electric
charge in a quantum ring is quantized in units of the electron charge
$e$ if the interaction is repulsive ($\alpha \leq 1$;
see  Fig.~\ref{fig:QofV.nospin}b). The step widths corresponding to
even ($\Delta V_{g}^{(e)}$) and odd ($\Delta V_{g}^{(o)}$) values of $\Delta N$
depend on the correlation parameter $\alpha$ and the dimensionless flux
$\Phi/\Phi_{0}$. The ratio of the widths  is described by the simple equation

\begin{equation}
\frac{\Delta V_{g}^{(p)}} {\Delta V_{g}^{(d)}} =
\frac{ \frac{1}{\alpha^{2}}-1+4 \frac{\Phi}{\Phi_{0}} }
     { \frac{1}{\alpha^{2}}+1-4 \frac{\Phi}{\Phi_{0}} } .
\label{deltaV}
\end{equation}
Recall that when deriving expression (\ref{meanN}) for the average number of
ring particles, the number $N_{0}$ of ``vacuum'' electrons was assumed to be
odd. Therefore the odd(even)-step corresponds to an even(odd) total number of
electrons in a ring.

\begin{figure}
\centerline{\psfig{figure=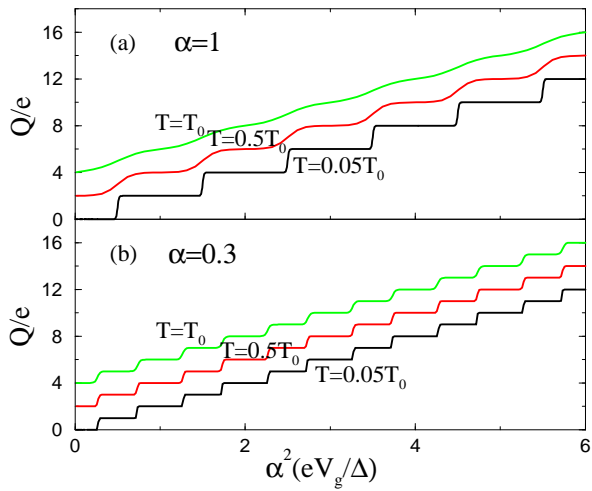,width=9cm}}
\caption{ \protect{\label{fig:QofV.nospin}}
Gate voltage dependence of the normalized charge for free (a) and
interacting (b) electrons without spin at different temperatures
(in units of $T_0=\Delta/2\pi$, where $\Delta$ is the energy level
spacing at the Fermi level). Traces have been vertically offset for clarity.
Note that the step width in (a) is the sum of the (unequal) step widths of two
adjacent steps in (b). Comparing a and b curves one can see that interactions
makes the Coulomb staircase more robust with respect to temperature.
}
\end{figure}

{}From Eq.~(\ref{deltaV}) one can see that in a stiff Wigner crystal-ring
($\alpha \ll 1$) all steps are of equal width --- as it should be -- since in
this limit the ground state energy does not depend on the parity of the
particle number. As $\alpha \rightarrow 1$, the steps corresponding to an even
total number of particles are reduced to a point and for the  LL with
attractive interactions ($\alpha >$ 1) the charge in the ring is quantized in
units of $2e$. The above picture of charge quantization (which is obvious from
the physical point of view) explains the oscillations between dia-
and  paramagnetic currents  analyzed in the previous
section. Notice that the ratio (\ref{deltaV}) of the step widths is
sensitive to
$\alpha$ in the region  $\alpha \leq 1$. This permits us to suggest the simple
experiment of measuring the charge quantization in a quantum ring as a
sensitive method of studying correlations in a Luttinger liquid.

At high temperatures, $T \agt T^{*}_{c}(\alpha)$, the quantization of charge
in a quantum ring is destroyed and the Coulomb staircase is smeared out into a
weakly modulated linear dependence of charge on gate voltage

\begin{eqnarray}
\Delta &&N(V_{g},T) = \frac{2\alpha^{2}}{\Delta} \left\{
eV_{g} - 4\pi T {\rm e}^{-4\pi^2\alpha^{2}T/\Delta} \sin(4\pi\alpha^{2}
\frac{eV_{g}}{\Delta}) + \right. \nonumber \\
&&
\left.
 4\pi T(-1)^{N_{0}} \cos(2\pi\frac{\Phi}{\Phi_{0}})
{\rm e}^{-\pi^2(1+\alpha^{2})T/\Delta} \sin(2\pi\alpha^{2}
\frac{eV_{g}}{\Delta}) \right\} .
\label{tworows}
\end{eqnarray}
It follows immediately that when $\alpha^{2} <$ 1/3,
the periodic modulation mainly comes from the second term of
Eq.~(\ref{tworows}).
 It depends neither on magnetic flux nor on the parity of ``vacuum'' electrons.
The crossover temperature for charge oscillations is equal to

\begin{equation}
T^{*}_{c} = \frac{\Delta}{4\pi^2 \alpha^{2}} , \quad \alpha^{2} < 1/3 .
\end{equation}
In the limit of strong repulsive interactions ($\alpha \rightarrow$ 0),
$T^{*}_{c} \rightarrow \infty$, which implies that a sharp
charge quantization is restored at any finite temperature.

Finally, when $\alpha^{2} >$ 1/3 the charge oscillations are determined by the
third term in Eq.~(\ref{tworows}) rather than the second.
The oscillation period
with respect to $V_{g}$ is then doubled  and the crossover temperature
decreases
with the increase of $\alpha$

\begin{equation}
T^{*}_{c} = \frac{\Delta}{\pi^2(1+\alpha^{2})} , \quad \alpha^{2} \geq 1/3 .
\end{equation}

\section{Influence of Long-Range Coulomb Interactions on the Thermodynamics
of Quantum Rings}

So far it has been assumed that the interactions between electrons in the
quantum ring is short ranged. Now we consider long-range Coulomb forces
and demonstrate that this interaction can be included in our scheme by
replacing the ``bare'' coupling constant $\alpha$ by an effective
size-dependent ``interaction constant'' $\alpha_{eff}$(L),
\begin{equation}
\alpha^{2}_{eff}(L) = \frac{\alpha^{2}}
{1+\alpha^{2} \frac{2}{\pi} \frac{e^{2}}{\hbar v_{F}} \ln
\left( \frac{L}{\lambda} \right)} .
\label{alphaeff}
\end{equation}
Here $\lambda$ is an ultraviolet cutoff ($\lambda \ll$ L) which in our case
has the physical meaning of the width of the quantum wire.

Long-range Coulomb interactions can be incorporated in our model by adding to
the local Lagrangian (\ref{Lagrangian}) a nonlocal Coulomb term
\begin{equation}
{\cal L}_{c} = -\frac{e^{2}}{8 \pi} \varphi^{'}(x) \int^{L}_{0} dy
\frac{ \varphi^{'}(y) }{ \sqrt{(x-y)^{2} + \lambda^{2}} } .
\label{LagrangianC}
\end{equation}
This term (see e.g. Refs.~\onlinecite{Schultz.prl,Fabrizio94}) describes the
electrostatic interaction of charge densities $\rho(x) = \varphi^{'}(x)/2\pi$
localized in a long 1D quantum wire of width $\lambda$.

In the presence of unscreened Coulomb forces, the equation of motion for
the dynamical field $\varphi(x,\tau)$ in imaginary time takes the form
\begin{equation}
\ddot{\varphi}+s^{2}\varphi^{\prime\prime} +
\frac{1}{2} s_{0}^{2} \frac{\partial}{\partial x} \left\{
\int^{L}_{0} \frac{\partial_{y} \varphi(y,\tau)}{\sqrt{(x-y)^{2} +
\lambda^{2}}} \right\} = 0 ,
\label{nonlocal}
\end{equation}
where $s_{0}^{2} = 2e^{2}\overline{\rho}/m_{0}$ ($\rho=m_{0}/L$).
Since we are interested only in topological excitations, we will seek
solutions to Eq.(\ref{nonlocal}) of the form
\begin{equation}
\varphi(x,\tau) = 2\pi \frac{\tau}{\beta} n + \psi(x), \quad n=0,\pm 1,...
\end{equation}
where the spatial derivative of $\psi$ ($\rho(x)=\partial_{x}\psi(x)/2\pi$)
obeys the integral equation
\begin{equation}
\rho(x) = \gamma \int^{L}_{0} \frac{dy \rho(y)}{\sqrt{(x-y)^{2} + \lambda^{2}}}
= C(\gamma) .
\label{C}
\end{equation}
Here $\gamma = (4 \alpha^{2}/\pi)(e^{2}/\hbar v_{F})$;
C($\gamma$) is an integration constant, which will be determined in what
follows by the requirement that the desired solution $\psi(x)$ has to satisfy
the twisted boundary conditions (\ref{boundary}).
One readily verifies that
\begin{equation}
\rho(\gamma) \simeq \frac{ C(\gamma)}{1+2\gamma \ln
\left( \frac{L}{\lambda} \right)}
\label{C2}
\end{equation}
is an approximate solution of the integral equation (\ref{C}).

It should be noted that when modelling the effects of long-range Coulomb
interactions in a quantum ring, we substitute the ring geometry by a
line when calculating the Coulomb energy, Eq.(\ref{LagrangianC}). Such a
substitution can be justified only for a long enough ring $L \gg \lambda$,
when one can neglect numerical factors in the argument of the big
logarithm in Eq.~(\ref{C2}). In this case the precise values of the integration
limits ($L \rightarrow aL$, $a\sim 1$) in the Coulomb integrals
Eqs.~(\ref{nonlocal}) and
(\ref{C}) do not affect the final expression evaluated in the
logarithmic approximation.
It is evident also from a physical point of view that in a perfect ring,
a homogenous distribution of charges in the long wavelength limit can be
inferred
from symmetry considerations.

By determining the integration constant C($\gamma$) from the requirement that
the
desired solution $\psi(x)$ satisfies the twisted boundary
conditions (\ref{boundary}),  one concludes that long-range Coulomb forces do
not deform the zero modes which in  the ideal (impurity free) ring remain of
the
the form of Eq.~(\ref{zeromodes}). However, the appearance  of a Coulomb energy
in the Lagrangian of the model changes the Eucledian action for these
trajecories. This effect can be taken into account by replacing the ``bare''
constant $\alpha$ by the  effective size-dependent coupling constant
$\alpha_{eff}(L)$ of Eq.~(\ref{alphaeff}).

Equation (\ref{alphaeff}) for $\alpha_{eff}(L)$ requires a comment. Let us
define the correlation parameter, $\alpha$, of the LL as the value of
$\alpha_{eff}$ at $L=\lambda$  ($\alpha \equiv \alpha_{eff}(\lambda)$). Then
Eq.~(\ref{alphaeff}) takes the typical form of a running interaction constant
in
the renormalization group sense (see e.g. Ref.~\onlinecite{Schultz}). As $L$
goes
to infinity,  $\alpha_{eff} \rightarrow$ 0 irrespective of the value of the
``bare'' coupling constant $\alpha$.
Hence, in the presence of long range Coulomb interactions, a LL-system of
long enough length behaves like a stiff ($\alpha_{eff} \ll 1$)  Wigner
crystal (see the discussion in Ref.~\onlinecite{Schultz.prl}).
In particular, the crossover temperature for current oscillations,

\begin{equation}
T_{j}^{*}(\alpha) = \frac{\Delta}{\pi^2}
\frac{ 1+\frac{2}{\pi} \alpha^{2} \frac{e^{2}}{\hbar v_{F}}
\ln \left( \frac{L}{\lambda} \right) } { 1+\alpha^{2}(1+\frac{2}{\pi}
\frac{e^{2}}{\hbar v_{F}} \ln \left( \frac{L}{\lambda} \right) ) } ,
\end{equation}
ceases to depend on $\alpha$ when
$(e^{2}/\hbar v_{F}) \ln \left( L/\lambda \right) \gg 1$ and coincides
with the crossover temperature of isolated LL-rings. \cite{Loss.prl,Krive}

In experiments involving quantum wires and dots in the 2D electron gas
of GaAlAs heterostructures, long-range Coulomb interactions can be partially
screened by a large metallic electrode (gate) which controls the density
of charge carriers. If the distance, $D$ ($D \gg \lambda$), between the quantum
ring and the gate is smaller than the characteristic size of the ring,
Coulomb forces are screened on distances of the order of $D$ and
one has to replace $L$ by $D$ in
Eq.~(\ref{alphaeff}). In the experiments considered,
the distance $D$ can be regarded as a controllable parameter. Thus one has
an interesting possibility to study oscillation effects in quantum rings
of different stiffness $\alpha_{eff}(D) < 1$.

\section{Thermodynamics of Quantum Rings for Spin-1/2 Electrons}
In the previous section we studied the thermodynamic properties of 1D
mesoscopic rings of strongly interacting spinless electrons.
Now we consider the influence of electron spin on the persistent
current and charge quantization in a quantum ring. It is well known that
in a perfect (impurity free) Luttinger liquid charge and spin degrees of
freedom are separated and their dynamics can be described by independent
quadratic Lagrangians (see e.g. Ref.~\onlinecite{Schultz}). Though the local
dynamics
of charge and spin excitations is independent, globally the two sectors of the
LL model are connected by the ``parity rules'' which reflect the simple
fact that the total spin and charge of the system are determined by the
numbers of spin-``up'' and spin-``down'' electrons. Therefore the spin
degrees of freedom affect persistent current- and
charge oscillations through the
parity term in the Lagrangian.

It is easy to generalize the model Lagrangian (\ref{Lagrangian})
to describe a Luttinger liquid of spin-1/2 electrons. Let
$N^{(0)}_{\uparrow,\downarrow}$ be the number of ``up''- and ``down''- spin
electrons in the ground state (T=0, $V_{g}$=0).
We will assume that the electron
system under consideration is not spin polarized on a macroscopic
scale. It means that the densities of electrons with
opposite spin projection are equal $\rho_{\uparrow} \simeq  \rho_{\downarrow} =
\overline{\rho}/2$, where  $\overline{\rho} =
(N^{(0)}_{\uparrow}+N^{(0)}_{\downarrow})/L$ is the mean electron density fixed
by the chemical potential $\mu$ of the electron reservoir. If we neglect the
difference in interaction of electrons with different spin projections, the
electron system in question can be regarded as a  two-component
Luttinger liquid of spinless electrons, \cite{Kusmartsev}

\begin{eqnarray}
{\cal L} &=&
\frac{m_{0}}{8\pi^2 \overline{\rho}} \sum_{j=\uparrow,\downarrow}
\left\{
\dot \varphi^2_{j} - \case{1}{4}
v_{F}^2(\varphi^\prime)^2 + \right. \nonumber \\ &&
\left. \frac{\hbar}{L} \dot\varphi_{j}
\left( \frac{\Phi}{\Phi_{0}} + \frac{N^{(0)}_{j}+m_{j}-1}{2} \right)
 + \frac{eV_{g}}{2\pi} \varphi^{'}_{j}
\right\}\nonumber \\ &&
-\frac{1}{8\pi^{2}} \sum_{i,j=\uparrow,\downarrow}
 \int dy U(x-y) \varphi^{'}_{i}(x)\varphi^{'}_{j}(y) .
\label{Lspin}
\end{eqnarray}

Here $v_F$ --- as above --- refers to the Fermi velocity of
spinless electrons of the same density.
The actual Fermi velocity here is therefore
$v_F/2 = \pi \hbar \overline{\rho}/2m_{0}$; U(x-y) is the kernel of the
electrostatic electron-electron interaction. In what follows we will assume
it to be local \cite{Kusmartsev}
\begin{equation}
U(x-y) = V_{0} \delta(x-y)
\end{equation}
The topological excitations which determine the current and charge oscillations
in the 2-component LL are of the form
\begin{equation}
\varphi_{j}(x,\tau) = n_{j} \frac{2\pi\tau}{\beta} + m_{j} \frac{2\pi x}{L},
\label{boundaryspin}
\end{equation}
where ($n_{j},m_{j}$) are winding numbers, which independently run over
integer values.
The Euclidean action for the trajectories (\ref{boundaryspin}) takes
the form
\begin{eqnarray}
S_{E}&&(n_{\uparrow,\downarrow},m_{\uparrow,\downarrow}) =
\sum_{j=\uparrow,\downarrow} \left\{
{2\pi^2} \frac{n^{2}_{j}}{\beta \Delta} + i2\pi n_{j}(\Theta_{j} +
\frac{m_{j}}{2})\right. \nonumber \\
&&\left. + \case{1}{8} \beta \Delta \left(1+4V_0/\pi\hbar v_F\right)
 m^{2}_{j} + \beta \mu_{g} m_{j}   \right\} \nonumber \\
&&
+\frac{V_{0} \beta}{L} m_{\uparrow} m_{\downarrow},
\end{eqnarray}
where
\begin{equation}
\Delta = \frac{2\pi\hbar v_{F}}{L} ,
\quad \Theta_{j} = \frac{\Phi}{\Phi_{0}} +
\frac{N^{(0)}-1}{2} , \quad \mu_{g} = eV_{g}
\end{equation}
When calculating the thermodynamic potential $\Omega$($\Phi,V_{g},T$) of a 1D
ring of strongly correlated electrons
\begin{equation}
\Omega-\Omega_{p} = -T \ln \left\{
\sum^{\infty}_{m_{\uparrow,\downarrow},n_{\uparrow,\downarrow}=-\infty}
\exp \left[ -S_{E}(n_{\uparrow,\downarrow},m_{\uparrow,\downarrow})/\hbar
\right]
\right\}
\label{Twoomegas2}
\end{equation}
it is convenient to replace the sum over
($m_{\uparrow,\downarrow},n_{\uparrow,\downarrow}$) by a sum over linear
combinations of winding numbers which describe the charge (c) and spin (s)
channels:
\begin{equation}
m_{c} = m_{\uparrow} + m_{\downarrow} ,  \quad
m_{s} = m_{\uparrow} - m_{\downarrow} .
\end{equation}
The summation in
Eq.(\ref{Twoomegas2})
can be performed ecactly. We omit the straightforward but cumbersome
intermediate calculations and  simply give the final expression for the
oscillating part of $\Omega$. The form of this expression depends significantly
on the parity of the numbers ($N^{(0)}_{\uparrow},N^{(0)}_{\downarrow}$) of
spin-''up'' and ``down'' vacuum electrons in the ring. Because of this we
separately analyze the cases  when the above numbers are of equal parity --
(odd,odd) and (even,even) or they have opposite parities (even,odd) or vice
versa. To compare the results for electrons with and without spin
we express all
quantites in terms of $I_{0}$ determined for the
spinless case:
\begin{equation}
I_{0} = \frac{e v_{F}}{L} = \frac{e \pi \hbar N^{(0)}}{m_{0}L^{2}} .
\label{spindef}
\end{equation}

\section{Quantum Ring with an Odd Number of Electron Pairs}
\label{sectionodd}
For the case of an odd number of electron pairs (spin-$\uparrow,\downarrow$),
the oscillating part of the thermodynamic potential $\Omega$ takes the form

\begin{eqnarray}
\label{oddodd}
\Omega^{osc}_{- -}&&= -4 \alpha^{2}_{c} \frac{(eV_{g})^{2}}{{\Delta}}
- \\
T \ln &&
\left\{
\theta^{2}_{3}(f,q^{2})
\left[ \theta_{3}(\overline{\mu},q^{\alpha^{2}_{c}}) \theta_{3}(0,q) +
\theta_{4}(\overline{\mu},q^{\alpha^{2}_{c}}) \theta_{4}(0,q) \right]
\right. \nonumber \\ &&
+ \theta^{2}_{4}(f,q^{2})
\left[ \theta_{3}(\overline{\mu},q^{\alpha^{2}_{c}}) \theta_{4}(0,q) +
\theta_{4}(\overline{\mu},q^{\alpha^{2}_{c}}) \theta_{3}(0,q) \right]
\nonumber \\ && \left.
+4 \theta_{4}(2f,q^{4}) \theta_{4}(2 \overline{\mu},q^{4\alpha^{2}_{c}})
\theta_{4}^{2}(0,q^{4}) \right\} ,
\nonumber
\end{eqnarray}
where $f \equiv \Phi/\Phi_{0}$ is the dimensionless flux,
$\overline{\mu} = 2\alpha^{2}_{c}eV_{g}/{\Delta}$; the parameters
$q$ and ${\Delta}$ are defined by
Eq.~(\ref{six}).
The correlation parameter $\alpha_{c} = v_{F}/v_{c}$ in the charge channel
--- where $v_{c}$ is the velocity of charged excitations --- is
\begin{equation}
\alpha^{2}_{c} = (1 + \frac{4 V_{0}}{\pi \hbar v_{F}})^{-1} ,
\label{alphac2}
\end{equation}
where
$v_{F} = \pi \hbar(N^{(0)}_{\uparrow}+N^{(0)}_{\downarrow})/m_{0}L$).
The parameter $\alpha_c$
describes the correlation properties of a LL of spin-1/2 electrons.
In the absence of any  Pauli interaction of the electron spin with the magnetic
field, the stiffness in the spin system coincides with that of a Fermi gas
of noninteracting particles, $\alpha_{s} \equiv v_{F}/v_{s}$ = 1.
These simple equations for the correlation parameters of the spin-1/2 LL
were first derived in Ref.~\onlinecite{Matveev,twentythree}.

To begin with we analyze Eq.~(\ref{oddodd}) in the limit of
noninteracting electrons ($V_{0} \rightarrow 0, \alpha_{c} \rightarrow 1$). In
this case  Eq.~(\ref{oddodd}) reduces to
\begin{eqnarray}
\label{free}
\Omega^{osc}_{- -} (\alpha_{c}&&=1) = -T \ln \left\{
2 \exp(2\overline{\mu})\right. \times \\ && \left.
\left[ \theta_{3}(f,q^{2}) \theta_{3}(\overline{\mu},q^{2}) +
\theta_{4}(f,q^{2}) \theta_{4}(\overline{\mu},q^{2}) \right] \right\} .
\nonumber
\end{eqnarray}
By making use of the addition formulae for the Jacobi theta-functions
(see e.g. Ref.~\onlinecite{Wittaker}) we get the following expression for the
persistent current of free spin-1/2 fermions at finite temperature
(Cf. Eqs.~(\ref{current1}) and (\ref{current3}))

\begin{eqnarray}
\label{free2}
I^{--}_{free}&& (\Phi,V_{g},T) = \frac{eT}{\hbar} \frac{1}{\pi}\times \\ &&
\frac{d}{df} \ln \left\{
\theta_{3}(f,q^{2}) \theta_{3}(\overline{\mu},q^{2}) +
\theta_{4}(f,q^{2}) \theta_{4}(\overline{\mu},q^{2}) \right\}
\nonumber \\ &&
= 4 \frac{eT}{\hbar} \sum^{\infty}_{n=1} (-1)^{n}
\frac{ \sin(2\pi n \frac{\Phi}{\Phi_{0}}) \cos(4\pi n \frac{eV_{g}}{\Delta})}
     { \sinh(4\pi^2 n \frac{T}{\Delta})}
\nonumber
\end{eqnarray}
This result coincides with the one
derived by Kulik \cite{Kulik70} in a
standard approach using the Fermi-Dirac distribution function $f_{FD}(p_{n})$,
\begin{eqnarray}
\label{fermidirac}
I(\Phi,\mu,T)&=& -c \sum_{n} \frac{\partial \epsilon_{n}}{\partial \Phi}
f_{FD}(p_{n}) \\ &=&
 4 \frac{eT}{\hbar} \sum^{\infty}_{k=1} \frac{
\sin(2\pi k \frac{\Phi}{\Phi_{0}}) \cos(4\pi k \frac{\mu}{\Delta}) }
{\sinh(k \frac{4\pi^2 T}{\Delta})},
\nonumber
\end{eqnarray}
where
$\epsilon$ is the one-particle spectrum. Since the persistent current is a
property of the Fermi surface, it does not depend on the precise form of the
spectrum far from the Fermi momentum. When expressed in terms of the Fermi
velocity, it is the same for quadratic and linear energy
dispersion. \cite{Cheung}

The standard expression (\ref{fermidirac}) for the persistent current in the
case of noninteracting electrons with spin is transformed to our
result (\ref{free2}) if we put  the chemical potential equal to $\mu =
\epsilon^{(0)}_{F} + eV_{g}$ ($\epsilon^{(0)}_{F}$ is the Fermi energy of the
ground state with an  odd number of spin-``up'',``down'' electron pairs). From
a physical point of view, it is evident that for noninteracting spin-1/2
electrons, the energy levels are doubly degenerate. Therefore the
ground state ($T=0$, $V_{g}$=0, $\Phi$=0) of free electrons at fixed chemical
potential has an odd number of electron pairs. We will prove this fact
explicitly below when studying the effects of the Coulomb blockade.

Notice that at $T=0$ and $V_{g}$=0, Eq.~(\ref{free}) describes the persistent
current for a system with a fixed number of particles. In the case considered
(odd number of pairs), our formula exactly coincides with the one obtained
in Ref.~\onlinecite{Kulik70}.

We conclude that the persistent current in a ring with an odd number of
spin-``up'' -``down'' pairs oscillates with the fundamental period $\Phi_{0} =
hc/e$. It is diamagnetic and at low temperatures has the same amplitude
$I_{0}=ev_{F}/L$ as the current of spinless fermions
(at the same particle density $\overline{\rho}$).
The influence of spin on the thermodynamic properties of quantum rings
of free electrons is straightforward --- the crossover temperature and
oscillation period on gate voltage are diminished by a factor 2
(in comparison with the spinless case, Eq.~(\ref{current3}))
which is a trivial consequence of the ``halving'' of the Fermi velocity
for electrons with spin and same density.

What new effects do the electron-electron correlations lead to?
At low temperatures ($T\rightarrow$ 0) the windings in the spatial sector are
frozen (due to the suppression of particle fluctuations) and we get a simple
expression for the persistent current of a LL-ring at $T=0$ of the form
\begin{equation}
I(T=0) = \frac{2}{\pi} I_{0} \sum^{\infty}_{n=1} (-1)^{n}
\frac{\sin(2\pi n \frac{\Phi}{\Phi_{0}})}{n}
\cos(4\pi n \alpha^{2}_{c} \frac{eV_{g}}{\Delta}) .
\label{fortyone}
\end{equation}
Although the oscillations have the same sawtooth-like shape as for
noninteracting particles, the period of the current oscillations when the
gate voltage is varied is drastically different in the case of strongly
correlated electrons. According to Eqs.~(\ref{alphac2}) and
(\ref{fortyone}) one gets for strong repulsion
current oscillations with a period
%
%
which is twice the
oscillation period for a LL-ring of spinless fermions
(see Fig.~\ref{fig:IofV}b).

Electron-electron correlations significantly affect the Aharonov-Bohm
oscillations (even for an impurity free ring) in the high temperature
regime. In particular, the crossover temperature becomes a function
of the correlation parameter $\alpha$. Using a small-$q$ expansion of the
$\theta$-functions, one can from Eq.~(\ref{oddodd}) derive the high-$T$
expansion
for the persistent current of correlated electrons. As in the case of
spinless fermions (see Eq.~(\ref{twoharm})) it is sufficient to keep
only two harmonics in the expansion,
\begin{eqnarray}
\nonumber
I_{--} \simeq&& -\frac{32}{3} \frac{eT}{\hbar}  \left\{
{\rm e}^{ -\pi^2(3+\alpha^{2}_{c})T/\Delta}
\cos \left( 4\pi \alpha^{2}_{c} \frac{eV_{g}}{\Delta} \right) \times
 \right. \\ &&
\sin \left( 2\pi\frac{\Phi}{\Phi_{0}} \right) +
\left. {\rm e}^{-8\pi^2 T/\Delta}
\sin \left( 4\pi\frac{\Phi}{\Phi_{0}} \right)
\right\} .
\label{fortytwo}
\end{eqnarray}
Note that the current
oscillations are determined mainly by the first term in (\ref{fortytwo}) if
$\alpha^{2}_{c} < 5$. The current is  diamagnetic and persists with fundamental
period $\Phi_{0}$. The crossover temperature is
\begin{equation}
T^{*}(\alpha^{2}_{c}) = \left\{
\begin{array}{cl}
\frac{\Delta}{\pi^2 (3+\alpha^{2}_{c})} \quad &
, \alpha^{2}_{c} < 5 \\
\frac{\Delta}{8\pi^2} &
, \alpha^{2}_{c} \geq 5
\end{array}
\right.
\label{fortythree}
\end{equation}

The transition to the regime where the oscillationperiod is reduced to
$\Phi_{0}/2$ occurs only when the interaction is strong and
attractive,  i.e. when $\alpha^{2}_{c} > 5$. However, it should be noted that
unlike in the spinless case, the two-component model (\ref{Lspin}) can not be
directly applied for describing attractive interactions. It is known (see
e.g. Ref.~\onlinecite{Schultz}) that in the last case backscattering processes
become relevant and result in a gap for the spin excitation spectrum.
The appearance of the gap rules out Luttinger liquid-like behavior
of spin-1/2 attractive electrons. Therefore, in what follows we will
analyze only the regime of repulsive interactions ($\alpha_{c} < 1$).

It is worth to note here that the inclusion of spin degrees of freedom changes
the maximum crossover temperature, which is attained in the limit of strong
repulsion $\alpha_{s} \ll$ 1 (stiff Wigner crystal). As one can see from
Eq.~(\ref{fortythree}), for spin-1/2 electrons this temperature is three
times smaller than for the Wigner crystal-ring of spinless fermions,
Eq.~(\ref{crossover}).

Now we proceed to the analysis of charge quantization in a LL-ring of electrons
with spin. For the case in question (an odd number of electron pairs:
$N^{(0)}_{\uparrow} = N^{(0)}_{\uparrow}$ =2n+1) the analytic formulae for
$\overline{N}(V_{g})$ are too cumbersome for practical use and we will consider
only numerical results. Some results for electrons with spin are shown in
Fig.~{\ref{fig:staircase}.

\begin{figure}
\centerline{\psfig{figure=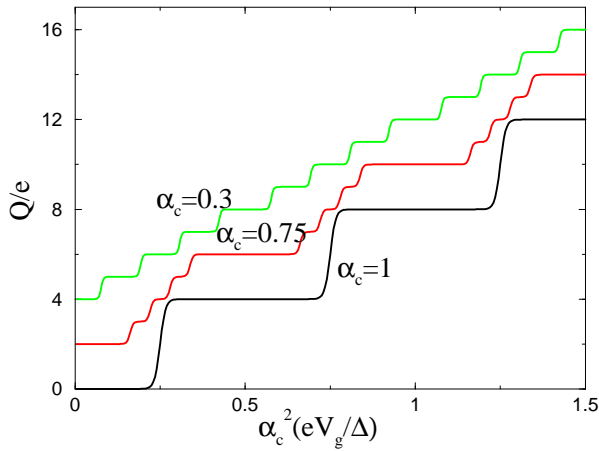,width=9cm}}
\caption{\protect{\label{fig:staircase}}
Splitting of the Coulomb staircase at low temperatures and no flux
due to repulsive interactions between
spin-1/2 electrons. Traces have been vertically offset for clarity.
In the absence of a magnetic flux the energy levels of the
noninteracting electrons ($\alpha_c=1$) are fourfold degenerate (with a factor
of two from spin).
Hence four electrons at a time are entering the ring each time
the gate voltage $V_g$ is such that a new degenerate ring level is pulled below
the chemical potential of the reservoir. As the strength of the repulsive
electron-electron interactions is increased ($\alpha_c < 1$) the
noninteracting energy levels become increasingly irrelevant and electron
correlations dominate. As shown this implies that electrons are entering the
ring one by one, and in the limit $\alpha_c\ll 1$ the steps in the Coulomb
staircase become equidistant. }
\end{figure}

In the strongly repulsive limit ($\alpha_{c} \ll 1$) and at low temperatures
the spin degree of freedom is insignificant and we have the same Coulomb
staircase as for spinless electrons.
The charge quantization for the two cases considered (electrons with and
without spin) differs drastically in the opposite limit --- weakly interacting
particles. For spin-1/2 fermions, the steps pertaining to the numbers N =
$4n$, $4n+1$, $4n+3$ electrons on the ring are decreased with the decrease  of
interaction strength and disappear (the width of the steps is contracted to a
point) at $\alpha_{c} \rightarrow 1$.

The above picture demonstrates the physical fact that at fixed chemical
chemical potential (grand canonical ensemble) the stable ground state of
the Fermi gas of spin-1/2 fermions always contains an odd number of
spin-``up'' and spin-``down'' electrons
$N^{(0)}_{\uparrow} = N^{(0)}_{\uparrow}$ = $2n+1$. It leads to a diamagnetic
response of a ring when applying an external magnetic field.

Since the width of the other steps goes to zero as
$\alpha_{c} \rightarrow$ 1, one can estimate the destruction temperature
of the corresponding ``vacua'' for $\alpha_{c} \alt$ 1,
$T_{d} \sim (1-\alpha^{2}_{c}) \Delta/2$
($\Delta/2$ is the level spacing for the Fermi gas of
electrons with spin).


\section{Quantum Ring with an Even Number of Electron Pairs}
The thermodynamic potential, $\Omega_{++}$, for an even number of
 pairs of ring electron with opposite spin projections
$N^{(0)}_{\uparrow} = N^{(0)}_{\uparrow}$ =2n  can be derived from
Eq.~(\ref{oddodd}) by a shift of flux, $f \rightarrow f$ + 1/2.
This is the parity rule for electrons with spin. \cite{Weisz}
In general, the thermodynamic potentials (or more correctly their oscillating
parts) for the two cases in question --- (odd,odd) and (even,even) ---
obey the following symmetry relations:
\begin{equation}
\Omega^{osc}_{++}(f,\overline{\mu}) = \Omega^{osc}_{--}(f+1/2,\overline{\mu})
= \Omega^{osc}_{--}(f,\overline{\mu}+1/2) .
\end{equation}
By shifting arguments in the expression (\ref{free}) valid
for an odd number of
noninteracting electrons we find the persistent current for  noninteracting
electrons in a ring with an even number of pairs to be
\begin{equation}
I^{++}_{free}(\Phi,V_{g},T) = 4 \frac{eT}{\hbar}
\sum_{n=1}^{\infty}
\frac{ \sin(2\pi n \frac{\Phi}{\Phi_{0}}) \cos(4\pi n \frac{eV_{g}}{\Delta})}
     { \sinh(4\pi^2n \frac{T}{\Delta})} .
\label{ee}
\end{equation}
This expression transforms into the standard result Eq.~(\ref{fermidirac}) for
$\mu = \epsilon^{(e)}_{F} + eV_{g}$, where $\epsilon^{e}_{F}$ is the Fermi
energy of the ground state with an {\em even} number of electron
pairs. As was discussed in section \ref{sectionodd}, such a ground state
($N^{(0)}_{\uparrow} = N^{(0)}_{\uparrow}$ =2n) for noninteracting
particles is destroyed due to particle exchange with the reservoir.
Thus Eq.(\ref{ee}) makes sense only in the T=0, $V_{g}$=0 limit:
\begin{equation}
I^{++}_{free}(T=0,V_{g}=0) = \frac{2}{\pi} I_{0} \sum_{n=1}^{\infty}
\frac{\sin(2\pi n \frac{\Phi}{\Phi_{0}})} {n},
\label{eveneven}
\end{equation}
when it describes the persistent current of a ring with fixed ($N=4n$) number
of particles. Eq.(\ref{eveneven}) coincides with the one obtained in
Ref.~\onlinecite{Loss.prb}. The current is paramagnetic, $\Phi_{0}$-periodic
and has the same amplitude, $I_{0}$ as the persistent current of spinless
fermions. The temperature behavior of the persistent current for a
LL-ring in contact with a reservoir is described be Eqs.~(\ref{fortytwo}) and
(\ref{fortythree}), independent of the parity of the number of
``vacuum'' electrons
($N^{(0)}_{\uparrow},N^{(0)}_{\uparrow}$).

\section{Quantum Ring with an Odd Number of Particles}
For a ring with an odd number of spin-1/2 fermions
($N^{(0)}$ = $4n+1$ or $4n+3$),
the oscillating part of the thermodynamic potential
$\Omega ^{osc}_{+-}$, takes the form

\begin{eqnarray}
\label{evenodd}
\Omega_{+-}&& = \Omega_{-+} = -T \ln
\left\{ \exp\left( 4\alpha^{2}_{c}
\frac{\mu^{2}_{g}\beta} {{\Delta}} \right)\right.\times \\ &&
\left[
\theta_{3}(2 \frac{\Phi}{\Phi_{0}}, q^{4})
\theta_{4}(4\alpha^{2}_{c}
\frac{\mu_{g}}{{\Delta}},q^{4\alpha^{2}_{c}}) + \right. \nonumber\\
&& \left.\left.
\theta_{4}(2 \frac{\Phi}{\Phi_{0}}, q^{4})
\theta_{3}(4\alpha^{2}_{c}
\frac{\mu_{g}}{{\Delta}},q^{4\alpha^{2}_{c}})
\right]
\right\}
\nonumber
\end{eqnarray}
The principal difference between Eqs.~(\ref{evenodd}) and (\ref{oddodd})
is concerned with the periodicity in flux and gate voltage. Using the
properties of the Jacobi theta-functions, one can immediately see
from Eq.~(\ref{evenodd}) that in a ring with an odd number of particles,
the persistent current is periodic in magnetic flux with period  $\Phi_{0}/2$,
half the fundamental period.

Period ``halving'' for the persistent current of an odd number of spin-1/2
noninteracting fermions was predicted (for $T=0$) in
Ref.~\onlinecite{Loss.prb}. From the general properties of Eq.~(\ref{evenodd})
we can conclude that this effect survives in the presence of interactions
(see also Ref.~\onlinecite{Yu,Fujimoto}) and even for finite temperatures.
However, Eq.~(\ref{evenodd})  pertains to the case of an odd number of
``ground state'' electrons in the ring. Therefore one has to be sure that
vacuum states corresponding to an odd number of particles ($N=4n+1$ or $4n+3$)
are stable against particle number fluctuations since we study the situation
with a fixed chemical potential.
We have already seen that it is not the case for noninteracting spin-1/2
fermions. In the last case, the ground state corresponds to
$N^{(0)} = 4n+2$ number of electrons.
Thus the $\alpha_{c} \rightarrow 1$ limit of Eq.~(\ref{evenodd}) can be
justified only for strictly zero temperature and $V_{g}$=0 when it
describes properties of a ring with fixed (odd) number of noninteracting
electrons. In this limit we get from Eq.~(\ref{evenodd}) a simple
expression for the persistent current,
\begin{equation}
I^{+-}_{free}(T=0,V_{g}=0) = \frac{1}{\pi} I_{0} \sum_{n=1}^{\infty}
\frac{\sin(4\pi n \frac{\Phi}{\Phi_{0}})} {n},
\label{evenoddI}
\end{equation}
which coincides exactly with the one in Ref.~\onlinecite{Loss.prb}.
The current is
paramagnetic and periodic with half the period, $\Phi_{0}/2$, and
half the amplitude, $\frac{1}{2} I_{0}$, in comparison with the case of
spinless electrons.

For repulsive interactions, the ``odd'' ground state is stable only at
temperatures $T \ll T_{d} \sim (1-\alpha^{2}_{c}) {\Delta}$.
Therefore it is reasonable to study current oscillations for the
(even-odd)-case only in the low temperature regime. The high temperature
behavior of the current is always described by Eq.~(\ref{fortytwo}).

At low temperatures the oscillations of current with varying gate voltage at
$\Phi \rightarrow$ 0 can be interpreted as oscillations between a dia-
and a paramagnetic response. Qualitatively these oscillations
are of the same form as for the spinless particles
(see Fig.~\ref{fig:IofV}).
For strong repulsion ($\alpha_{c} \ll$ 1, stiff Wigner crystal) the bumps
of paramagnetic response and dips of diamagnetic response have the same
width in gate voltage. This property is closely related to the effects of
charge
quantazation. For a stiff Wigner crystal, the ``even'' and ``odd'' steps of the
function $e\Delta N(V_{g},T \rightarrow 0)$ =
$- (\partial \Omega_{+-} / \partial V_{g})$
have equal widths since in this strongly repulsive limit considered
the energy does not depend on the parity of the electron number.
With the increase of $\alpha_{c}$ the steps corresponding to an odd total
number of electrons on a ring get narrow and paramagnetic bursts of
current reduce to spikes. In the region  $\alpha_{c} \agt$ 1  (even,odd),
the ground state becomes absolutely unstable and to study the persistent
current of noninteracting electrons at fixed chemical potential, one
has to start from a different expression for $\Omega$, namely
Eq.~(\ref{free}), ``built'' on the stable ground state.

This work was supported by the Royal Swedish Academy of Science (KVA), the
Swedish Natural Science Research Council (NFR), INTAS Grant No. 94-3862 and by
Grant No. U2K200 from the Joint Fund of the Government of Ukraine and the
International Science Foundation. One of us (I.K.) acknowledges the hospitality
of the Department of Applied Physics, CTH/GU.


\begin{references}
\bibitem{note1}
The only exception is in the regime of very
strong magnetic field, when  under FQHE conditions it was shown \cite{Milliken}
that edge states do behave as a chiral LL. \cite{Wen}
\bibitem{note2}
The sum in Eq.~(\ref{potential})
was presented in a general form in Ref.~\onlinecite{Loss.prl}
(for $\mu_{g}$=0), however the summation has been done only for the case
of low temperatures. The final expression derived in
Ref.~\onlinecite{Loss.prl} describes the persistent current of an isolated
LL-ring (fixed number of particles) at finite temperatures and the low-T
limit for the case of a fixed chemical potential
(see also Ref.~\onlinecite{Krive}).
\end{references}
\end{document}